# 3D Printing: Developing Countries Perspectives

Fredrick R. Ishengoma
Computer Engineering and Applications
The University of Dodoma
P.O. Box 490, Dodoma, Tanzania

Adam B. Mtaho
Computer Engineering and Applications
The University of Dodoma
P.O. Box 490, Dodoma, Tanzania

## ABSTRACT
For the past decade, 3D printing (3DP) has become popular due to availability of low-cost 3D printers such as RepRap and Fab@Home; and better software, which offers a broad range of manufacturing platform that enables users to create customizable products. 3DP offers everybody with the power to convert a digital design into a three dimensional physical object. While the application of 3DP in developing countries is still at an early stage, the technology application promises vast solutions to existing problems. This paper presents a critical review of the current state of art of 3DP with a particular focus on developing countries. Moreover, it discusses the challenges, opportunities and future insights of 3DP in developing countries. This paper will serve as a basis for discussion and further research on this area.

## General Terms
3D Printing and Developing Countries

## Keywords
Additive Manufacturing and Rapid Prototyping.

## 1. INTRODUCTION
3D printing (3DP) (also known as Additive Manufacturing (AM) [1,2,3], Rapid Prototyping (RP) [4,5] or Stereo Lithography (SL) [6]) is a process of creating 3 dimensional physical objects from a 3D computer models made in Computer Aided Designs/Manufacturing/Engineering (CAD/CAM/CAE). The process is accomplished by using additive process. An additive process uses various materials; from biodegradable plastic filament polylactic acid (PLA) to Acrylonitrile Butadiene Styrene (ABS) plastic to nylon, whereby materials are melted into thin layers and layer upon layer are printed until the desired physical object is obtained. 3DP technology can be used to print almost anything from product prototypes to spare parts for machinery and appliances to printing jewelry.

In developed countries, 3DP has already taken off and is applied in diverse sectors including architecture, construction, healthcare, manufacturing, engineering, clothing, eyewear and education. Recently, the technology is gaining popularity in developing countries. The potential of 3DP to developing countries is huge. Envision individuals in a small remote village in Africa who could use 3DP technology to print basic and advanced products (like farming tools and bicycle spare parts) to improve their livelihood and productivity. Entrepreneurs in developing countries could design and manufacture items that without 3DP would need large upfront investing.

3DP presents an enormous opportunity to improve community livelihoods in developing countries by letting the local manufacturing firms (in urban and rural) design and produce innovative, robust and cost-effective products that can overcome the existing deficit in manufacturing value chain and support their lives and increase income.

This paper explores the state of art of 3DP in developing countries, and analyzes its opportunities and challenges. Moreover, the paper provides the future insights of 3DP application in developing countries. The case studies presented in this study are from the following developing countries: Uganda, South Sudan, Colombia, India, South Africa, Kenya, Trinidad and Tobago and Haiti.

The rest of the paper is structured as follows: Section 2 presents the evolution of 3DP. Section 3 describes how 3DP works. Section 4 discuses the existing case studies of 3DP in developing countries. Opportunities of 3DP in developing countries are discussed in Section 5. Section 6 discusses challenges that are facing developing countries in the adoption and the use of 3DP technology. We present our future insights on 3DP in developing countries in Section 7. We conclude in Section 8.

## 2. EVOLUTION OF 3D PRINTING
Charles Hull first engineered 3D printing technology in 1984 and called it Stereo Lithography (SL). He acquired a patent for the technology later in 1986. Several of his patents are still used in today's additive manufacturing processes. In the early years, 3D printers were very expensive and not made for general market. While SL systems were becoming popular by the end of 1980s, other comparable technologies such as Selective Laser Sintering (SLS) [7] and Fused Deposition Modeling (FDM) [8,9] emerged.

Massachusetts Institute of Technology (MIT) engineered another technology named "3 Dimensional Techniques" in 1993 [10] and attained a patent for the technology licensing it to Z corporation [11]. The technology uses an inkjet technology similar to that used in 2D printers. Z Cooperation launched a first high definition (HD) 3D color printer named Spectrum Z510 in 2005.

Another advancement in 3DP happened in 2006 with the start of an open source project, named RepRap project [12]. The project aimed at creating a self-replicating 3D printer that has capability to print most of its own making objects. RepRap project uses Fused Filament Fabrication (FFF) to place down material in layers. RepRap uses ABS, Polylactic Acid (PLA), nylon, HDPE and similar thermo polymers to print objects. The project is an open source project whereby all projects are released under a free software license, the GNU General Public License [13]. As we entered into the 21st century, the costs of 3D printers significantly fell, allowing 3D printers to be affordable to the general market [14].

## 3. HOW 3D PRINTING WORKS

3D printing is made by printing layer after layer forming the physical model with the reference to the digital 3D model designed by 3D CAD software or scanned from a 3D scanner. At first, a design is modeled using CAD software in a computer, and then a layer of printing material is fed into the 3D printer. The print head is used to apply the printing material (example resin) in the shape of the model. Layers are added until the desired physical model is achieved. The sequence of printing is shown in figure 1.

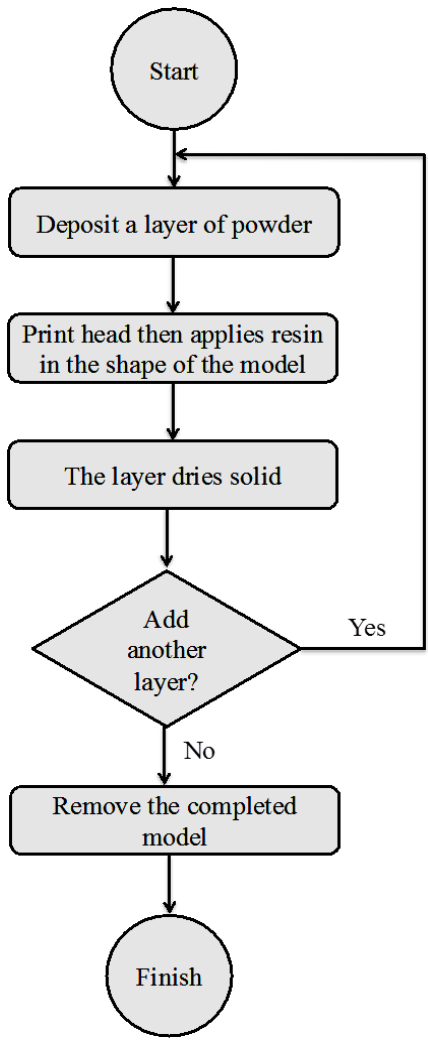

**Fig 1: 3D printing sequence**

3DP is applied in different fields such as in industrial designs, automotive design, and consumer commodities, engineering, dental and aerospace. Traditionally, large companies have been applying the technology to create prototypes before the final production. Recently, the technology has been applied in full-scale manufacturing products.

## 4. APPLICATION OF 3D PRINTING IN DEVELOPING COUNTRIES

Currently most 3D printers present in developing countries are limited to universities under engineering departments and dedicated workshops such as Fabrication Laboratories (FabLabs) [15]. FabLabs are small-scale kind of factory equipped with industrial-grade fabrication, electronics tools and softwares used by individuals to manufacture prototypes. FabLabs are increasingly being adopted in developing countries whereby currently they are available in South Africa, India, Kenya, Ethiopia, Ghana, Burkina Faso, Senegal, Colombia, Peru, Argentina, Chile, Costa Rica, Mexico, Malaysia, Indonesia, Pakistan, Thailand and Trinidad and Tobago.

Developing countries are faced with diverse health issues that are prevalent, amputation being one of them. In Uganda and South Sudan, 3DP technology is used to create 3D-printed prosthetic limbs. Due to lack of enough specialized prosthetic technicians, the scanned image of a patient is sent to different parts of the world where it is designed as a digital 3D model, and the file is sent back again for printing the prosthetic limb. 3D limbs are now used as an alternative of handmade limbs for some cases. Although 3D limbs don't have better quality like handmade limbs, most patients prefer them because fitting of 3D limbs is not time consuming compared to traditional limbs. 3DP therefore provides a faster and cheaper approach to deal with prosthetic situations in developing countries than traditional limbs. The components used in 3D printing of prosthetic limbs include fiber and 3D molds, which make the prosthesis to be lightweight, replaceable and flexible. In South Africa, a company called Robohand [16] is manufacturing prosthetic hand, arms and fingers using 3DP technology and thermoplastics. In developing countries where healthcare services strive to keep up with its demand, and the high costs of medical care, 3D printing might well be a saving grace.

In Haiti, citizens and aids workers are still struggling to recover from the effects of 2010 earthquake. iLab Haiti [17] makes use of 3D printing to create prototypes for future printing of certain basic medical supplies used on demand, examples, umbilical clamps, finger splints and casts. The designed prototypes would be 3D printed for straight use in local clinics with on-demand manufacturing. Moreover, 3DP technology has been applied in Haiti to print shoes for school children in need.

In Kenyan FabLab, a 3D printed infant vein finder was developed. The prototype enables doctor to find veins in infants patients and manage important intravenous needles to infants, which is usually a challenging task. Meanwhile, researchers in the University of Trinidad and Tobago are creating anatomical models using 3DP [18]. The technology enables the surgeon and healthcare institutions to explore the benefits of using 3DP in healthcare services. The research group is working to advance the application of 3DP technology in healthcare services. One of its future major aims is to use 3DP to conduct pre-operative planning for surgeons resulting in faster surgical processes.

In Colombia, 3DP have been applied in innovative way in fashion designing industry making new innovative fashion dress and accessories. In India, 3DP have been applied to empower waste pickers by buying waste plastics at a reasonable cost. The plastics are then converted to printer filament. This reduces the cost of 3DP service in the local area.

## 5. CHALLENGES

In this section we discuss various challenges facing 3DP technology in developing countries.

### 5.1 Availability of 3DP Experts

To fully deploy the use of 3D printing, developing countries need to have enough number of experts who can design, model and manufacture different products. So far there is a shortage of experts who can design and manufacture different products using 3D printing technology. More training are needed in developing countries so as to increase the availability experts of 3DP technology.

### 5.2 Copyright in 3DP
Another challenge that may arise is the copyright issue. Copyright in 3DP refers to unregistered right that binds the protection of the created original 3D models. Most of the 3D CAD models are shared online in the form of open source whereby anyone can download the model and print the physical object. Recently, the traditional players in manufacturing and engineering industry have been going after websites that share the 3d models online reasoning that home printers are tearing their intellectual property. 3DP could face legal challenges in digital rights management (DRM) to inhibit online file sharing. Copyright laws in developing countries so far have not implemented the law that governs sharing and use of 3D models online. However, as 3DP becomes more common, the challenge of printing copyrighted materials will increase and become nearly difficult to determine.

### 5.3 Decrease in Manufacturing Jobs
3DP presents a platform for individuals to print models and things even without the knowledge of making them; this will increase the number of manufacturing jobs, since people now have the ability to print mostly anything in their homes. This disadvantage can impact the economies of developing countries.

### 5.4 Limited and Fixed Materials
Currently, the 3D printers create products by making use of plastics, resin, ceramic and metals. The common 3D Printers are still not capable of mixing different materials required for printing. As a result, for a product that needs a mixture of more than one material as constituents, its quality or strength is affected. For example, printing models that contains both rubber and metals is still challenging. Meanwhile the large commercial 3D printers that support mixed materials are very expensive for majority of users from developing countries.

### 5.5 Size
The common 3D printers currently available in the market are still limited for printing small sized objects. Therefore, this limits printing large equipment such as huge plastic water tanks (with more than 20,000 liters capacity). With the advancement of technology, in the near future 3DP technology is envisioned to be used to print large objects such as bricks for low cost houses.

## 6. OPPORTUNITIES
In this section we discuss various opportunities that arise in developing countries following the use of 3DP technology.

### 6.1 Economic Empowerment
Developing countries are regularly detached from world supply chains for even basic commodities; nonetheless 3D printing has the capability to bring developing countries into the loop. 3DP is guaranteed to give everybody in the developing countries the power to manufacture or just create virtually whatsoever for their own uses. For example, developing countries can use 3DP technology to manufacture local equipment such as toys, farming tools, domestic tool etc. This will help create new jobs and empower people economically.

### 6.2 Improving Science Education
Science and technology are the important aspects of economic development in the developed countries. Developing countries also needs science and technology to improve the lives of the citizens. Henceforth with this 3DP technology, it is anticipated that the teaching of science in developing countries will advance. In the developed countries, 3D printing is basically used by students to generate models/prototypes of things deprived of the use of costly tooling needed in subtractive methods. Students design and manufacture physical models. The classroom permits students to study and engage innovative applications for 3DP. This can be replicated to developing countries.

### 6.3 Improving Science Education
3DP can be used to assist developing countries to conserve the diminishing forest cover because of careless logging. Just for example, if furniture and many additional domestic things would be made from plastic, few trees would be cut and several more conserved. Henceforth developing countries will have donated a great deal in addressing the adverse global effects of climate change, for example global warming.

### 6.4 Emergence Response
Technology has a great potential to make a huge difference in emerging responses in developing countries by printing needed materials like basic tools rather than flying them from another countries. Imagine if following an earthquake citizen can use 3DP to construct houses within a day. What if citizens had inexpensive and available medical kits that could create bespoke medicine on demand? Envisage if citizens could construct shoes, lamps, pumps, clothes and just about whatsoever else on demand. These examples show how 3DP can be used to develop tools that can assist in responding quickly to several forms of emergencies.

### 6.5 3D Printed Agricultural Tools
3DP technology presents an opportunity for farmers in developing countries to create agricultural tools on demand hence increase their economy. Currently, most of small farmers are using labour intensive agricultural hand tools. However, due to the increase in rural to urban migration in developing countries, particularly among youth [19, 20], this human labour force is decreasing day after day. Innovating less-intensive labour agricultural hand tools can be designed and developed for agriculture and household food processing to improve livelihood and meet daily needs. This includes tools for harvesting, weeding, planting and food processing. Blacksmiths and artisan in developing countries could use 3DP technology to develop innovative prototypes to meet agricultural system requirements and community at large. 3DP technology could be used by local farmers and blacksmiths to develop both income and agricultural products. It is predicted that the rise in adoption of 3DP in developing countries will reduce the dependence on Chinese and Indian imports.

### 6.6 Surgical Application
Recently survey studies [21, 22] have shown that, developing countries have nearly zero access to vital surgical services because of lack of access to facilities or basic equipment. With the remarkable adoption of 3DP technology worldwide, as the cost of 3D printers falls, and access to electricity and Internet improves in developing countries, the problem of surgical services will be solved. For example, currently Fripp [23], a UK based company is printing 150 false eyes, ears and noses for developing countries. This mass production practice considerably reduces the production cost. The mass production technique could be even more cost-effective if the production is made in-house (in a particular developing country). With advancement of Internet access in developing countries, availability of 3DP experts and knowledge sharing, it is possible the developing countries to be printing complex models like skull and skin for the patients in the near future.

### 6.7 Low Energy Solutions
The process of creating renewable energy sources to produce electricity in developing countries as a credible solution is a challenge. Various innovative concepts of using 3DP for renewable energy are being tested. One of the concepts that employed 3DP was applied during the Haiti 2010 earthquake to generate emergency power in the affected areas. The research is still on progress; the positive results of the research will have a major impact in the application of renewable energy in developing countries.

### 6.8 Water and Sanitation
The 3DP technology can be used to print rainwater catchment equipment using recycled plastics in developing countries. Additionally, the technology can be used to create "fresh water maker" equipment that can be used to generate fresh water from dirty water. The equipment heats the dirty water up to 85º C where water vapor starts to be generated. Water vapor is collected in the walls of the funnel and the condensed water flows to accumulating pool creating suitable and safe distilled water for drinking. This practice will assist in solving water and sanitation challenges. Also, 3DP technology could be used to print household items like soap holders and toilet items at low cost using recycled plastics.

### 6.9 Decrease Carbon Footprint
If many businesses decide to accept and implement 3DP practices, carbon emissions will decrease significantly. Since 3D-printed commodities can be created and composed in just one process, thousands of parts wouldn't have to be transported from overseas to developing countries. Furthermore, many product design files can also be transferred digitally and then printed on demand in developing countries, easing the necessity to find and construct space for inventories.

## 7. FUTURE
The future of 3DP in developing countries is very promising. In the near future, 3DP is expected to be applied in diverse sectors of our daily lives. 3DP could be used to print advanced medical objects like skin or plastic skull for advanced transplants. Envision 3D printer printing a skin directly to a patient in a war zone/battlefield in developing countries using skin cells, enzymes and collagen. 3DP could be used to print houses during disasters occurrence like Haiti 2010 earthquake. The clay building material could be fed in 3D printer to print out the desired bricks for making a house. Additionally, we envision 3DP used to print solar cells for emergence power and renewable energy in remote locations of developing countries.

## 8. CONCLUSIONS
3DP technology presents a huge prospect for developing countries. The technology offers local societies the ability to innovate, design and create tools that support and improve their daily lives. The technology can be used for economic empowerment and improving the livelihood of communities. However, the existing challenges such as availability of experts and small size of the available printers will be solved with time. Moreover, as technological advancement and Internet access improved in developing countries, 3DP technology is expected to revolutionize manufacturing sector worldwide.